# Characterization and shaping of the time-frequency Schmidt mode spectrum of bright twin beams generated in gas-filled hollow-core PCF

M. A. Finger,[1,*] N. Y. Joly,[2,1] P. St.J. Russell,[1,2] and M. V. Chekhova[1,2,3]

[1]*Max Planck Institute for the Science of Light, and* [2]*Department of Physics, University of Erlangen-Nuremberg, Staudtstraße 2, 91058 Erlangen, Germany*

[3]*M. V. Lomonosov Moscow State University, 119899 Moscow, Russia*

*Corresponding author: martin.finger@mpl.mpg.de*

We vary the time-frequency mode structure of ultrafast pulse-pumped modulational instability (MI) twin beams in an argon-filled hollow-core kagomé-style PCF by adjusting the pressure, pump pulse chirp, fiber length and parametric gain. Compared to solid-core systems, the pressure dependent dispersion landscape brings increased flexibility to the tailoring of frequency correlations and we demonstrate that the pump pulse chirp can be used to tune the joint spectrum of femtosecond-pumped $\chi^{(3)}$ sources. We also characterize the resulting mode content, not only by measuring the multimode second-order correlation function g[(2)] but also by directly reconstructing the shapes and weights of time-frequency Schmidt (TFS) modes. We show that the number of modes directly influences the shot-to-shot pulse-energy and spectral-shape fluctuations in MI. Using this approach we control and monitor the number of TFS modes within the range from 1.3 to 4 using only a single fiber.

## I. Introduction

The study of frequency correlations is an active research topic in both classical and quantum optics [1 - 7]. Recently, there has been increasing interest in investigating frequency correlations in the classical nonlinear optics domain, e.g., in supercontinuum generation [2, 3] and in modulational instability (MI) [1, 8]. This has applications in source development, generation of random numbers [9] and can be used to study other analogous nonlinear physical systems [1]. All experiments in the classical domain have so far only investigated frequency correlations but did not attempt to *modify* these correlations systematically. In contrast, in quantum optics, one of the challenges is source *engineering* [4, 5, 10, 11] and controlling photon correlations in all degrees of freedom (e.g., momentum, frequency, and polarization) is of fundamental importance for biphoton and twin-beam sources. However, the dependence of frequency correlations on the dispersion makes them difficult to modify in most systems [4, 10, 11].

In quantum optics, frequency correlations and entanglement are described by introducing time-frequency Schmidt (TFS) modes [5, 6, 12 - 15]. These are natural collective, broadband frequency modes of bipartite systems and a particular example of the broader concept of coherent modes, used in the classical description of partially coherent light sources [16, 17]. The remarkable feature of TFS modes is that, when signal and idler modes of twin beams are properly selected, their photon numbers are perfectly correlated. This deep physical meaning makes TFS modes a very useful tool in quantum optics [12]. However, their controlled generation, manipulation and detection is, despite many promising experiments, still a challenge [13, 15].

Engineered frequency correlations are highly relevant for many applications and concepts in quantum optics. Indeed, a moderate number of frequency modes is interesting for increasing the security of quantum key distribution [18] through using larger alphabets, or for enlarging the information capacity of optical communication channels [5, 6, 19]. Meanwhile, heralding multiphoton nonclassical states and high interference contrast experiments require ideally single-TFS-mode states [20, 21]. In the classical domain, shaping frequency correlations of MI is important because it allows one to reduce pulse-energy fluctuations. This is appealing for supercontinuum generation, since MI is often the initial broadening stage and reducing its pulse-energy fluctuations might lead to more stable supercontinuum sources. Apart from that, shaping frequency correlations makes it possible to change MI pulse-energy fluctuation statistics from an exponential distribution for a single TFS mode towards a Gaussian distribution for a large number of modes [22]. This leads to non-deterministic physical random number generators, where the underlying probability distribution can be changed in a controlled way. Such generators at optical frequencies and high repetition rates are interesting for modern information technology [9, 23]. This shows that engineering the frequency correlations is essential in both quantum optics and classical nonlinear optics. Since MI is a ubiquitous process in nonlinear wave propagation and has been observed in many different systems, from plasma physics to hydrodynamics, we expect that our work will also influence other fields of physics [24].

In this paper, we investigate the frequency correlations for twin beams generated by femtosecond pulse-pumped modulation instability in gas-filled kagomé-style photonic crystal fibers (PCFs), which are known as a workhorse for dispersion-tailored nonlinear optics [25]. We first introduce a novel and fast way to retrieve the TFS modes from joint spectral measurements. After that we demonstrate the tuning of the number of modes in three ways. One of them is changing the gas pressure, a unique feature of this system, which greatly simplifies the engineering of spectral correlations. The other one is applying a frequency chirp to the pump pulses. Previous work on biphoton states has shown that a pump pulse chirp alters the phase of the joint spectral amplitude and creates phase entanglement [26 - 28]. Here we show that such a chirp can also, through interplay with self-phase modulation, modify the joint spectral intensity, thus offering a new means of continuously varying the spectral photon-number correlations between signal and idler beams in all ultrafast pulse-pumped systems based on the $\chi^{(3)}$ nonlinearity. Finally, we investigate for the first time how the number of TFS modes changes with the parametric gain in a system without spectral filtering of the sidebands. We also observe how TFS modes manifest themselves in the shot-to-shot fluctuations of the pulse energy and the spectral shape.

In our previous work we showed that the spectral location of the photon-number correlated MI sidebands can be tuned [29]. Now, we show a more important and unique feature: that the modal frequency content is tunable and can be precisely adjusted on demand.

## II. Experimental setup and theoretical background

Our experiment focuses on the high-gain modulational instability in an argon-filled hollow-core PCF pumped by femtosecond pulses. In fiber optics, four-wave mixing (FWM) and modulational instability (MI) are often used as two names for the same nonlinear process, depending on whether the system is pumped in the normal (FWM) or anomalous (MI) dispersion regime. Pumping in the normal dispersion regime leads to narrow sidebands with a large separation from the pump, and requires pumping with long pulses to limit the effect of temporal walk-off. On the other hand, pumping in the anomalous dispersion regime generates broad sidebands located closely to the pump.

Figure 1 shows the experimental set-up. We pump our system with 140 fs pulses from an amplified Ti:sapphire laser operating at 800 nm with a repetition rate of 125 kHz. The pumped fiber is a kagomé-style hollow-core PCF (kagomé PCF) with a core diameter of 18.5 μm and a core wall thickness of 240 nm. It is filled with argon at a pressure of ~75 bar yielding a zero dispersion wavelength at 770 nm, so that the pump is in the anomalous dispersion regime. The pulse chirp is controlled using a single-prism pulse compressor, based on a highly dispersive SF66 prism and a retroreflector [30]. It allows us to introduce a chirp of up to -40,000 fs$^2$. After the compressor, we use a bandpass filter (BP) to reduce the input spectral width. All experiments with unchirped pulses are carried out with a 3.1 nm BP (transform-limited pulse duration: 280 fs) in order to reduce the self-phase modulation (SPM) broadening of the pump, while the experiments with chirped pulses employ a 5 nm BP (transform-limited pulse duration: 240 fs). The pulses are characterized using a commercial pulseCheck FROG from APE.

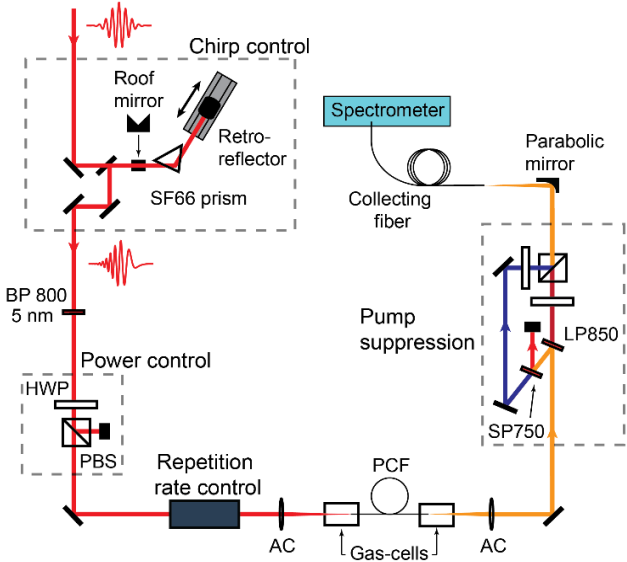

FIG. 1. (Color online) Experimental set-up for studying single-shot, chirp-dependent frequency correlations. The chirp is controlled using a single-prism compressor [30]. HWP – Half-wave plate, PBS – Polarizing beam splitter, AC – Achromat, BP – Bandpass filter, SP – Shortpass filter, LP – Longpass filter.

We detect the signal and idler spectrum after the Ar-filled kagomé PCF with a spectrometer (OceanOptics, Maya2000 PRO) with a minimum integration time of 6 ms. To avoid saturation of our detection we must remove the remaining pump, which can be as wide as 100 nm due to its SPM broadening. For this, we use a combination of a shortpass filter at 750 nm and a longpass filter at 850 nm to form a notch filter with a very broad bandwidth (Fig. 1). We take single-shot measurements to study directly the spectral correlations between sidebands. This is possible by reducing the repetition rate of our laser from 125 kHz down to 80 Hz by means of a Pockels cell placed between two orthogonal polarizers. This system has a comparatively high dynamic range of 40 dB for single-shot detection. Alternatively, we use photodiodes (not shown in the figure) to measure the signal or idler pulse energies with higher sensitivity and at the full repetition rate of 125 kHz. This is useful to quickly obtain the number of frequency modes in our system by measuring the normalized second-order autocorrelation function $g^{(2)} = \langle N_{s,i}^2 \rangle / \langle N_{s,i} \rangle^2$, where $N_{s,i}$ is the photon number in the signal or idler beams. As previously shown, g$^{(2)}$ is directly linked to the number of TFS modes in the system and is a measure for pulse-energy fluctuations [29].

In order to understand how frequency correlations are generated and how they can be influenced, it is instructive to look at the time-integrated Hamiltonian describing the quantum state of biphotons [14]:

$$\hat{H} \propto \iint d\omega_i d\omega_s F(\omega_i, \omega_s) \hat{a}^\dagger(\omega_i) \hat{a}^\dagger(\omega_s) + \text{h.c.} \quad (1)$$

In this expression $\hat{a}^\dagger(\omega_{s,i})$ are monochromatic-wave signal and idler photon creation operators and the joint spectral amplitude (JSA) $F(\omega_i, \omega_s)$ describes the frequency correlations between them. In the case of MI in a fiber of length $L$ pumped by a single source with the spectral amplitude $\alpha(\omega)$, the JSA is [10]

$$F(\omega_i, \omega_s) = \int d\omega' \alpha(\omega') \alpha(\omega_s + \omega_i - \omega') \\ \times \text{sinc}\left[\frac{L}{2} \Delta k(\omega_s, \omega_i, \omega')\right] \exp\left[i \frac{L}{2} \Delta k(\omega_s, \omega_i, \omega')\right], \quad (2)$$

where $\Delta k = k(\omega') + k(\omega_s + \omega_i - \omega') - k(\omega_s) - k(\omega_i) - 2\gamma P_p$ is the phase mismatch. $\gamma$ is the nonlinear parameter, which depends on the used fiber and the frequency and $P_p$ is the peak power [31].

This description is only strictly valid for biphotons, where time-ordering of the evolution operator can be neglected [32, 33]. Nevertheless, it is also very useful in the high-gain regime for understanding how best to tune the spectral correlations. The JSA is a complex-valued function and has for the first time only recently been experimentally reconstructed through phase-sensitive amplification [34]. In all previous experiments the joint spectral intensity (JSI = |JSA|$^2$) was measured either by spectrally resolved coincidence counting [35] or by stimulated emission tomography [36]. Fig. 2 shows the JSI numerically calculated for a 30 cm long kagomé PCF filled with 70 bar argon and pumped at 800 nm. The pump spectral amplitude is assumed to be Gaussian and only linear phase mismatch is considered [35, 10].

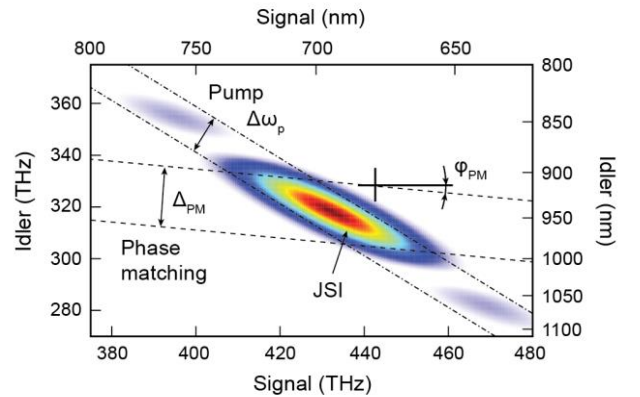

FIG. 2. (Color online) Joint spectral intensity (JSI) of a gas-filled kagomé-style photonic crystal fiber pumped in the MI regime at 800 nm, calculated with only linear phase mismatch taken into account. Δω$_P$ is the spectral width of the pump, $\Delta_{PM}$ is the full width at half maximum and $\varphi_{PM}$ the angle of the phase matching contour (given by the sinc- function in Eq. 2) with the horizontal axis.

With a known JSA, the TFS mode structure can be revealed by finding its Schmidt decomposition [12]. This involves expressing the JSA as an expansion in terms of two sets of independent orthonormal modes $\phi_n(\omega_s)$ and $\chi_n(\omega_i)$, each one depending on just the signal or the idler frequency, so that $F(\omega_i, \omega_s) = \sum_n \sqrt{\lambda_n} \phi_n(\omega_s) \chi_n(\omega_i)$. Here, $\lambda_n$ are known as the Schmidt eigenvalues with $\sum_n \lambda_n = 1$ and their amplitude decreases with increasing *n*. Physically, they mean the average photon number distribution over different modes. A certain system is called single-mode in frequency if the JSA can be written as a single product of some $\phi_0(\omega_s)$ and $\chi_0(\omega_i)$ modes, meaning in other words that the JSA is factorable. In this way, the orientation of the JSA – and of spectral cross correlations in general – is connected to the number of TFS modes and gives a first indication of how many modes exist in the system.

At a certain parametric gain *G*, which is proportional to the fiber length, nonlinearity and the pump power, the number of photons in each mode is [37]

$$N_n = \sinh^2\left(G\sqrt{\lambda_n}\right). \quad (3)$$

This shows that the contribution of an individual mode *n* to the total photon number changes with the parametric gain. Therefore, it is useful to introduce new (high-gain) Schmidt eigenvalues $\tilde{\lambda}_n = N_n / \sum_n N_n$, which reduce in the case of low gain (*G* << 1) to the normal Schmidt eigenvalues, $\tilde{\lambda}_n \to \lambda_n$ [37].

## III. Determining TFS modes from joint measurements

In the case of two-photon light, the JSI is found by measuring the coincidence rate between two detectors counting photons at various pairs of frequencies. For twin beams generated at high gain, such a coincidence measurement is impossible. Here, we quantify spectral cross-correlations at high gain by recording a set of single-shot spectra and calculating the covariance between the photon numbers registered at individual frequencies (Fig. 3(a), (b)) [38]. The covariance is defined as

$$\mathrm{Cov}\left[N(\omega_i), N(\omega_j)\right] = \langle N(\omega_i)N(\omega_j)\rangle - \langle N(\omega_i)\rangle\langle N(\omega_j)\rangle,$$

where $N(\omega_{i,j})$ is the photon number at frequency $\omega_{i,j}$ and the angle brackets denote averaging over different pulses. It turns out that the covariance can be expressed in terms of TFS modes and their eigenvalues (for the derivation, see Appendix A):

$$\mathrm{Cov}\left[N(\omega_s), N(\omega_i)\right] = \left|\sum_n^\infty v_n u_n \phi_n(\omega_s)\chi_n(\omega_i)\right|^2, \quad (4)$$

where $v_n = \sinh(G\sqrt{\lambda_n})$ and $u_n = \cosh(G\sqrt{\lambda_n})$. Thus, by taking the square-root of the measured covariance we get direct access to the shapes of the TFS modes. Since the covariance measurement is phase-insensitive, reconstructing the modes requires the assumption that the phase of the JSA is flat. The validity of this can be challenged on the basis of a subsequent $g^{(2)}$ measurement. Even when the phase cannot be neglected, the extracted effective number of TFS modes gives a lower bound for the number of modes excited in the system. Although asymmetric losses lead to distortions of the mode shapes, the approximate modal shapes can now be found by making a singular value decomposition of the square-root of the covariance. One then obtains the correct Schmidt eigenvalues by taking the square-root of the calculated singular values because in the high-gain regime $v_n \cong u_n$. We want to point out that in the low-gain regime the normalized covariance is exactly the JSI because $u_n = 1$ and $v_n \propto \sqrt{\lambda_n}$. In this sense, the covariance measurement is a novel way to retrieve the JSI, in addition to the coincidence measurement and stimulated emission tomography.

Figure 3 shows the TFS modes reconstructed for a fiber length of 0.3 m, a pressure of 76 bar and an unchirped 280 fs pulse (Fig. 3(a)). The covariance is calculated from 2500 single-shot spectra. Fig. 3(d) show the shapes of the first three signal and idler TFS modes. We can see that the mode shapes resemble Hermite-Gaussian functions, which are the analytical solutions for a double-Gaussian JSA [14]. Fig. 3(c) shows the distribution of the high-gain Schmidt eigenvalues (modal weights) for two cases. The blue bars correspond to Fig. 3(a) and the red bars to the nearly spatiotemporal single mode case of Fig. 3(b) (1.5 m, 71 bar and 390 fs pulses with a down-chirp of 29,000 fs$^2$). Clearly, fewer modes are excited in the case of the red bars. The effective number of modes at high gain *K* can be found from the Schmidt eigenvalues: $K = \left(\sum_n \tilde{\lambda}_n^2\right)^{-1}$. In the two cases shown in Fig. 3, *K* = 1.3 and 2.7.

The effective number of modes *K* is also related to $g^{(2)}$ via $g^{(2)} = 1 + 1/K$, which assumes that a single mode has thermal statistics [39], i.e., its normalized correlation function is $g^{(2)} = 2$ [29]. These relations can now be used to find $g^{(2)}$ directly from the shape of the covariance. At the same time, such a measurement gives much more information about the frequency mode structure than a normal $g^{(2)}$ measurement, which is currently the standard method to retrieve the number of TFS modes. We obtain good agreement between the $g^{(2)}$ value found by integrating the single-shot measurements ($g^{(2)}$ = 1.33±0.04) and the value found by analyzing only the spectral correlations ($g^{(2)}$ = 1.37±0.03, Fig. 3(c)). This confirms that the assumption of a flat phase is valid and justifies the reconstruction of the modal shapes from a phase-insensitive covariance measurement. To our knowledge, this is the first experimental demonstration of the intriguing link between frequency correlations and intensity fluctuations in MI. Note that in contrast with Fig. 2 (low gain theory), we cannot observe any side lobes in our measurement, and we attribute this to the high gain values [32] and the loss [40] in the system.

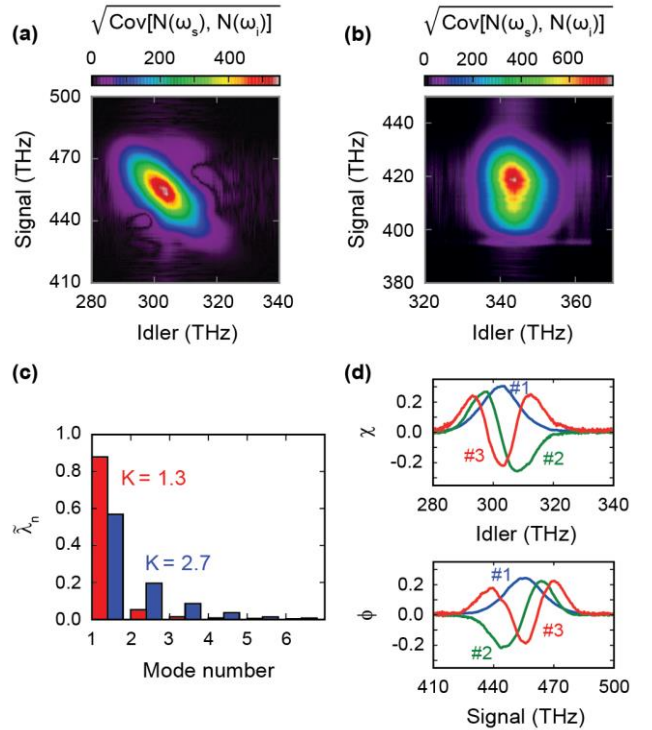

FIG. 3. (Color online) Square-root of the covariance in Eq. 4 for (a) 0.3 m, 76 bar and unchirped 280 fs pulses and (b) 1.5 m, 71 bar and 390 fs pulses, down-chirped at 29,000 fs$^2$. (c) Schmidt eigenvalues for the case shown in (a) (blue bars), resulting in *K* = 2.7, and in (b) (red bars) giving *K* = 1.3. Panels in (d) show the first three TFS modes for the case presented in (a).

## IV. Shaping spectral correlations

The only possibilities to modify the JSI of solid-core fibers and crystals is so far to change the pump bandwidth, the interaction length or the nonlinear material [37, 4]. Here we show additional possibilities to fine tune frequency correlations, namely: changing the pressure, valid for gas-filled hollow-core fibers, and pre-chirping the pump pulses, valid for all ultrafast $\chi^{(3)}$-based systems.

### A. Pressure tuning

Pressure tuning allows one to balance the normal dispersion of the gas against the weak anomalous geometrical dispersion of the waveguide and thus change the dispersion landscape [41] and consequently also the MI spectrum [29]. As a result, the tilt angle $\varphi_{PM}$ of the phase-matching condition in Fig. 2 is no longer a constant as for solid-core fibers but becomes a function of the gas pressure. The tilt $\varphi_{PM}$ is determined by the group-velocity mismatch between signal, idler and pump photons and is given by $\varphi_{PM} = -\arctan\left[(\beta_{1,s} - \beta_{1,p})/(\beta_{1,i} - \beta_{1,p})\right]$ with respect to the signal axis [10]. Here, $\beta_{1,\mu} = \left[\partial\beta(\omega)/\partial\omega\right]_{\omega=\omega_\mu}$ with $\mu = s, i, p$ and $\omega_\mu$ are perfectly phase-matched frequencies and $\beta(\omega)$ is the propagation constant. Fig. 4 shows the dependence of the group velocity mismatch for the PCF under study, applying various argon pressures. While the mismatch between the idler and the pump is always positive, it can change sign on the signal side depending on the gas pressure. This is associated with a change in sign of the tilt angle $\varphi_{PM}$ and illustrates the pressure-dependent tunability of the phase matching.

We study the frequency cross-correlations of the generated sidebands by plotting the correlation coefficient $C_{ij}$. This is a normalized version of the covariance used above (preceding Eq. 4) and has the advantage that the strength of the correlations can be directly seen and easily interpreted, since $C_{ij}$ takes values between ±1. In particular, $C_{ij} = -1, 0, +1$ mean, respectively, perfect anti-correlation, no correlation and perfect correlation. The correlation coefficient is defined as $C_{ij} = \mathrm{Cov}\left[N(\omega_i), N(\omega_j)\right] / \sqrt{\Delta N(\omega_i)^2 \Delta N(\omega_j)^2}$, where $\Delta N(\omega_{i,j})^2$ represents the variance of the photon numbers at the frequencies $\omega_{i,j}$. $C_{ij}$ is symmetric with respect to frequency, so that $C_{ij} = C_{ji}$ (see Appendix B).

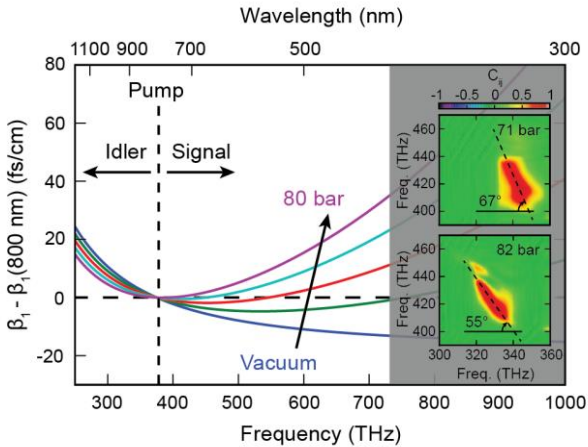

FIG. 4. (Color online) Group-velocity mismatch relative to the pump at 800 nm for various Ar pressures. Insets: Pressure-dependent change of the frequency cross-correlations in a 1.5 m long fiber pumped with 390 fs pulses, which had a down-chirp of 29,000 fs². $K = 2.5$ at 82 bar and 1.4 at 71 bar.

The inset of Fig. 4 illustrates the effect on the frequency cross-correlations of changing the pressure from 82 to 71 bar. The measurements were taken in a 1.5 m long PCF, pumped with 390 fs long pulses (down-chirped at 29,000 fs²). The tilt of the cross-correlation distribution can be seen to change from 55° to 67°, showing that by varying only the pressure, the frequency correlations can be modified. This changes the effective number of modes from $K = 2.5$ to 1.4.

### B. Chirp tuning

It is known that chirped pump pulses act on the JSA so as to introduce additional entanglement in the phase of the generated state [26 - 28]. Here we show that in a fs-pulse-pumped system the JSI also depends, through self-phase modulation, on the chirp of the pump pulses. This provides an additional means of tuning not only the phase correlations but also the spectral photon-number correlations in $\chi^{(3)}$-based sources. In ultrafast pulse-pumped systems, self-phase modulation (SPM) broadening of the pump is inherent to sources based on the Kerr-effect. Although usually considered disturbing, this effect can also be used to fine tune spectral correlations by pre-chirping the pump pulses. In the time domain, SPM broadening of a Gaussian pulse is to a good approximation associated with a linear frequency up-chirp. This is caused by a temporally varying nonlinear phase, which implies that the instantaneous frequency varies across the pulse, leading to the generation of new frequencies and spectral broadening. The SPM-induced chirp adds to the initial chirp of the pump pulses. Therefore, it is possible to control the effective pump width $\Delta\omega_p$ in the fiber by pre-chirping the input pulse [31]. Ultrafast pulse-pumped systems therefore offer an additional degree of freedom for controlling the spectral width $\Delta\omega_p$ (Fig. 2), which directly influences the frequency correlations. The chirp has the advantages of being continuously tunable and free of losses, unlike the filtering of the pump spectrum. It has been pointed out in [28] that it is difficult to obtain high rates of spectrally pure biphotons due to the effects of SPM. Here we report that pre-chirping the pump pulses can remedy this situation by compensating for SPM.

We investigate the influence of pre-chirping a 240 fs pump pulse on the frequency cross-correlation in a 1.5 m long PCF filled with 76 bar argon. From the top panel of Fig. 5(a) we see that with increasing chirp we can limit the SPM broadening in the fiber. This is on the one hand due to the compensation of SPM-induced chirp and on the other hand due to the reduced pulse peak power of the temporal-stretched pre-chirped pulse, since we keep the pulse energy constant. In this way, we can avoid the characteristic oscillations in the SPM spectrum. As we see in Fig. 5(a) (bottom right), a strongly SPM-broadened pump leads to distortion of the cross-correlation signal. However, if we chirp the input pulse more strongly so that the pump maintains a single-peak spectrum, we obtain a clean cross-correlation signal (Fig. 5(a), bottom left). In this way, chirping the input pulse allows one to obtain clean frequency correlations even in systems with strong, normally unavoidable, SPM.

Fig. 5(b) shows that the effective number of modes $K$ increases with increasing pre-chirp; in the figure it varies between ~1.7 and ~2.6. This behavior results from the chirp-induced reshaping of the joint spectrum and can be clearly distinguished from the previously reported creation of phase entanglement through the pump chirp [27]. In our system, this effect would lead to the opposite behavior (less modes with increasing pre-chirp) because of the progressive phase flattening due to the increasing compensation of SPM-induced chirp and pump pre-chirp. Here, we balanced the temporal-stretching of the pulse with an increase in pulse energy, so that the gain in the system was about constant. Despite the increase in pulse energy, the SPM broadening is still reduced because of the pre-chirp. In conclusion, varying the pump chirp allows the effective number of TFS modes in ultrafast pulse pumped systems to be continuously tuned. The exact functional form and magnitude depends on the phase matching (Fig. 2).

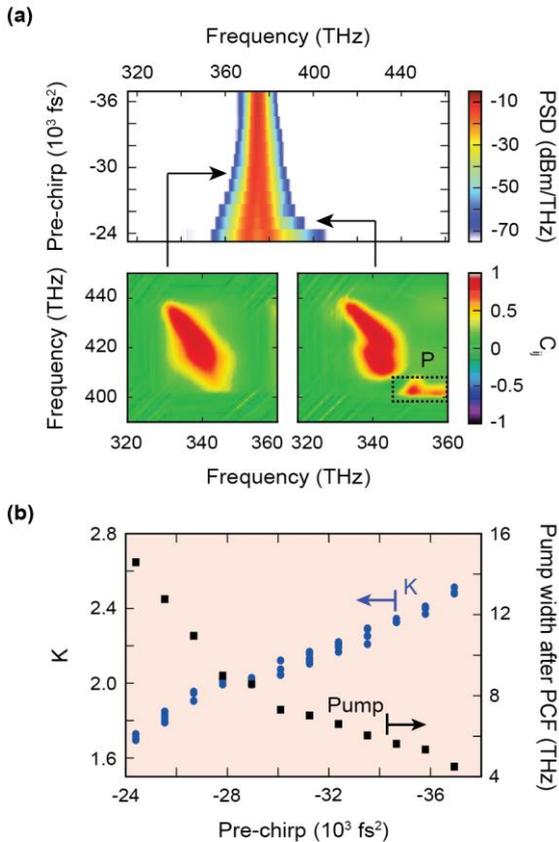

FIG. 5. (Color online) (a) Top: Influence of SPM on spectral cross-correlation. SPM-broadening is adjusted by varying the chirp of a pulse with constant energy (Average input power: 10 mW), propagating along 1.5 m PCF filled with 76 bar argon. Bottom left: Cross-correlation when the pump spectrum exhibits only a single peak. Bottom right: Distorted correlation for a strongly SPM-structured pump. The dashed rectangle marks the residual pump (P). (b) Effective mode number $K$ and spectral width of the pump after the fiber at $1/e^2$ of the pump amplitude versus the pre-chirp of the pump pulses. We balance the pulse duration increase caused by the pre-chirp with more pump power.

## V. Spatiotemporal few-mode source

In this section we show that, using the methods above, the number of TFS modes can be changed from 1.3 to 4 with a single fiber. Further, we investigate how the number of TFS modes influences the shot-to-shot spectra and the pulse-energy fluctuations and demonstrate that the number of TFS modes is not a constant but depends on the gain in the system.

### A. Influence of TFS mode number on shot-to-shot fluctuations

As we have seen, by changing the kagomé PCF length, the gas-filling pressure and the chirp of the input pulse the MI frequency correlations can be shaped. Fig. 6 compares the cross-correlations generated in two different configurations. The first system is a 30 cm long PCF, filled with 76 bar argon and pumped with unchirped 280 fs pulses (Fig. 6(a)-(c)). The second system is a 1.5 m long PCF at 71 bar, pumped with down-chirped (-29,000 fs$^2$) 390 fs pulses (Fig. 6(d)-(f)). In the first case (Fig. 6(a)), one can see that the high signal frequencies are correlated with the low idler frequencies and vice versa (frequency anti-correlation). The narrow width of the correlated region means that this anti-correlation is strong. In contrast, in the second case (Fig. 6(d)), the shape reveals almost no frequency correlations or anti-correlations. This difference is also reflected in the measured g$^{(2)}$ values, which change from g$^{(2)}$ = 1.33 in the first case to 1.71 in the second case. This means that pulse-energy fluctuations are higher whenever there are fewer frequency modes. Figs. 6(b) and (e) show 100 individually normalized single-shot spectra for the two cases. One can clearly see that the location of the peak intensity is strongly fluctuating in (b) and many slices exhibit more than one peak. In contrast, Fig. 6(e) shows that the spectral peak jitters much less and is mainly localized at the center. Moreover, most spectra exhibit now only a single peak. This effect is clearly visible when comparing Fig. 6(c) and (f) and is a consequence of the different number of TFS modes for the two systems. The shape of the spectra fluctuates strongly from shot to shot for $K$ = 3.0, which is the result of the superposition of differently populated, independent spectral modes. In contrast, for $K$ = 1.4 the shape of the spectra remains much more constant while only its amplitude fluctuates—an indication that a single mode is predominantly populated. The same reasoning also explains the higher pulse-energy fluctuations in the case of fewer TFS modes. The intrinsic pulse-energy fluctuations arise from the noise-seeded nature of the MI light and have thermal statistics (g$^{(2)}$ = 2) in the single-mode case, while becoming damped for multimode light. This figure illustrates neatly the applicability of the TFS mode picture to classical systems. Consequently, if the aim in a classical system is to reduce the shot-to-shot power fluctuations due to MI, it is beneficial to excite a large number of TFS modes.

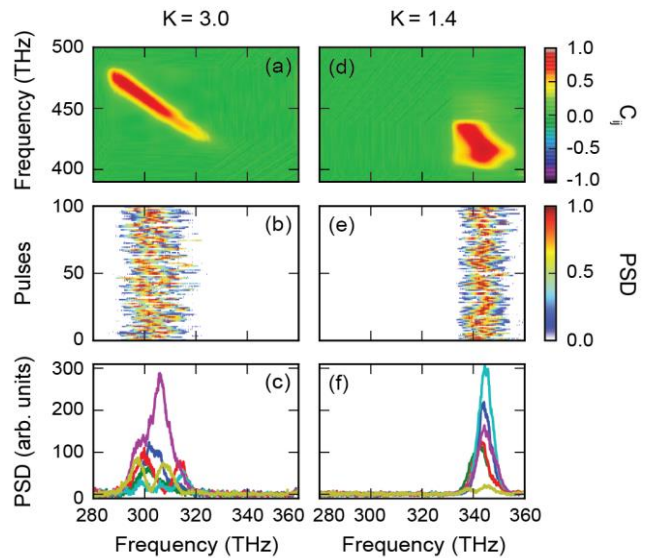

FIG. 6. (Color online) Influence of frequency correlations on spectral fluctuations in single-shot spectra. One can see that strong frequency correlations (i.e. a large number of TFS modes) lead to shot-to-shot fluctuations of the spectral shape, whereas a more factorable state leads to a more stable shape. Here, the parameters of the two systems are: 0.3 m, 76 bar, unchirped 280 fs pulses (left panels) and 1.5 m, 71 bar, 390 fs pulses (down-chirped with -29,000 fs$^2$) (right panels). (a), (d) Frequency cross-correlations. (b), (e) Power spectral density (PSD) of 100 individually normalized single-shot spectra. (c), (f) Plot of six consecutive single-shot spectra.

### B. Gain-dependent number of TFS modes

Finally, we made a direct measurement of $K$ as a function of pump power up to very large parametric gain values without spectral filtering. Theory predicts that the number of photons per mode scales with the parametric gain $G$ according to Eq. 3 [37]. Therefore, lower-order modes, which have larger eigenvalues (compare Fig. 3(c)), are more strongly amplified with increasing gain than higher-order modes. This can be interpreted as a gain-dependent redistribution of the normalized Schmidt eigenvalues. In the end, this theory predicts that every system collapses to a single frequency mode when the gain is high enough. Figure 7(a) shows gain-dependent measurements of $K$ for three different parameter sets with a photodiode: (I) 1.5 m, 71 bar, 390 fs pulses chirped with -29,000 fs$^2$; (II) 1 m, 76 bar, unchirped 280 fs pulses; and (III) 0.3 m, 76 bar, unchirped 280 fs pulses. One can see that the number of modes in the system reduces when the gain is increased, which confirms theoretical predictions [37]. Since the number of modes is equal to the product of the numbers of spatial and temporal modes, we can readily access the number of spatial modes by filtering out a narrow spectral band (4 nm) from the single-shot spectra (so as to obtain a single frequency mode). Fig. 7(a) shows the $g^{(2)}$ for a single frequency mode found in this way. We can see that the curve approaches 2 at moderate gain values, proving that the system is spatially single-mode. However, for higher gain even the filtered $g^{(2)}$ decreases, suggesting that the single-mode statistics change at high values of gain. This can be explained by two-photon loss of the signal in the fiber: at high gain, secondary sidebands form through pairwise escape of photons from the signal sideband. This effect, which resembles two-photon absorption, is known to suppress intensity fluctuations and can even lead to sub-Poissonian statistics [42, 43]. Indeed, by examining the gain-dependent spectral evolution of the sidebands (Fig. 7(b)), we observe that the reduction in $g^{(2)}$ begins when the secondary sidebands appear. This is indicated in Fig. 7(a) by the red arrow corresponding to the spectrum in red in Fig. 7(b). Subsequently, they start to spectrally overlap with the main sidebands, eventually resulting in the whole spectrum broadening to a supercontinuum. At this point spectral correlations vanish almost completely (see Appendix C).

Consequently, as the gain increases (Fig. 7(a)), we observe a reduction in the number of modes but cannot reach a value of $K = 1$ by just increasing the power because of the two-photon loss.

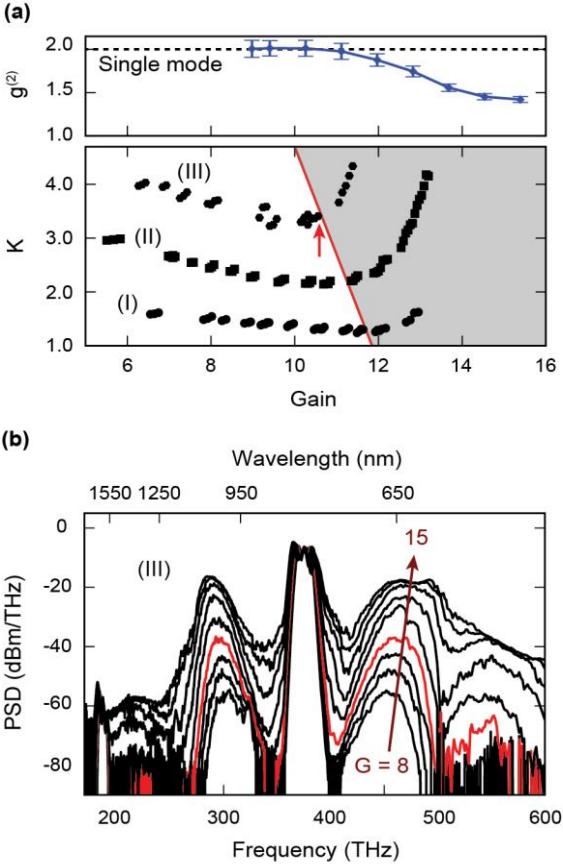

FIG. 7. (Color online) (a) Dependence of the effective mode number $K$ on the gain, which is varied by changing the pump power and measured with a photodiode. The single-mode $g^{(2)}$ is found by numerically truncating spectrometer measurements (II) to a 4 nm band. The error bars represent the standard error. The gray-shaded area marks gain values above which the single-mode statistics change. (I) 1.5 m, 71 bar, 390 fs pulses chirped with -29,000 fs$^2$; (II) 1 m, 76 bar, 280 fs pulses; (III) 0.3m, 76 bar, 280 fs pulses. The red arrow marks the gain value of the red trace in Fig. (b). (b) Spectral evolution of (III) when the gain (G) is increased from 8 to 15.

## VI. Conclusion

Bright twin beams with a tunable frequency mode content can be produced by modulational instability in gas-filled hollow-core kagomé PCF. Tuning the chirp of ultrafast pump pulses allows the JSI in Kerr-effect based systems to be precisely adjusted without introducing losses and distortions through truncating the pump spectrum and makes it possible to compensate SPM effects.

A unique feature of kagomé PCF is its flexibility in tuning spectral correlations by changing the pressure or the filling gas, something that is impossible in solid-core waveguides because of the rigid dispersion landscape. This simplifies source engineering and brings increased freedom to the experimental modification of spectral correlations, allowing observation of the connection between spectral correlations and single-shot fluctuations of MI spectral shape.

The shapes and weights of TFS modes can be retrieved from photon-number covariance measurements, which yield more information on the time-frequency mode structure than the standard $g^{(2)}$ measurements and enable experimental verification of theoretical predictions about TFS modes in the high-gain regime. This method can be extended to reconstructing the shapes and weights of coherent temporal and spatial modes in classical partially coherent radiation. Finally direct (unfiltered) measurements of the effective number of modes up to very high parametric gain values confirm that, in accordance with theory, the number of TFS modes in a system is not a constant but depends on the gain.

These studies contribute to the understanding and tailoring of frequency correlations, which is already of high importance for applications in quantum optics and will become also increasingly important in classical nonlinear optics, promising more stable supercontinuum generation or tunable physical random number generators.

## Appendix A: Derivation of the covariance for a twin-beam parametric amplifier in terms of TFS modes

We describe a spectrally multimode twin-beam parametric amplifier and derive its non-normalized second-order correlation function and covariance in terms of time-frequency Schmidt (TFS) modes. This description is valid for parametric down-conversion and four-wave mixing processes without pump depletion. It is known that the Schmidt modes form a basis for which the Hamiltonian is diagonal [12]. We express the negative-frequency field operators for signal and idler as a series expansion in terms of TFS modes,

$$\hat{E}_s^{(-)}(\omega_s) = \sum_{n=0}^{\infty} \hat{a}_n^{\dagger} \phi_n(\omega_s),$$
$$\hat{E}_i^{(-)}(\omega_i) = \sum_{n=0}^{\infty} \hat{b}_n^{\dagger} \chi_n(\omega_i).$$
(A1)

Here, $\phi_n$ and $\chi_n$ are the TFS modes of the signal and idler. These modes are orthogonal, so that $\int \phi_n^*(\omega)\phi_m(\omega)d\omega = \delta_{nm}$ and $\int \chi_n^*(\omega)\chi_m(\omega)d\omega = \delta_{nm}$. $\hat{a}_n^{\dagger}$ and $\hat{b}_n^{\dagger}$ are photon creation operators for the respective Schmidt modes, defined as

$$\hat{a}_n^{\dagger} = \int d\omega_s \hat{a}^{\dagger}(\omega_s)\phi_n(\omega_s),$$
$$\hat{b}_n^{\dagger} = \int d\omega_i \hat{b}^{\dagger}(\omega_i)\chi_n(\omega_i),$$
(A2)

where $\hat{a}^{\dagger}(\omega_s)$ and $\hat{b}^{\dagger}(\omega_i)$ are the standard monochromatic-wave photon creation operators. For the broadband operators the usual commutation relations hold, $[\hat{a}_n, \hat{a}_m^{\dagger}] = \delta_{nm}$, $[\hat{b}_n, \hat{b}_m^{\dagger}] = \delta_{nm}$, and $[\hat{a}_n, \hat{b}_m^{\dagger}] = 0$. The two-mode Hamiltonian is now of the form

$$\hat{H} = i\hbar \sum_n \Gamma_n \hat{a}_n^\dagger \hat{b}_n^\dagger + \text{h.c.} \quad (A3)$$

where $\Gamma_n = \Gamma\sqrt{\lambda_n}$ is the interaction strength, $\Gamma$ being the coupling strength for the whole amplifier and $\lambda_n$ the low-gain ("two-photon") Schmidt eigenvalues. For a parametric amplifier, the Bogoliubov transformations (input-output equations) always couple the TFS modes pairwise,

$$\hat{a}_n = u_n \hat{a}_{n0} + v_n \hat{b}_{n0}^\dagger,$$
$$\hat{b}_n = u_n \hat{b}_{n0} + v_n \hat{a}_{n0}^\dagger, \quad (A4)$$

where

$$u_n = \cosh\left(\int dt\, \Gamma_n\right) = \cosh\left(G\sqrt{\lambda_n}\right),$$
$$v_n = \sinh\left(G\sqrt{\lambda_n}\right). \quad (A5)$$

Here, $G \equiv \int dt\, \Gamma(t)$ is the parametric gain. From this, we can calculate the mean photon number in the signal channel,

$$\langle N(\omega_s)\rangle = \langle \hat{E}_s^{(-)}(\omega_s)\hat{E}_s^{(+)}(\omega_s)\rangle$$
$$= \sum_{n,k=0}^{\infty} \langle \hat{a}_n^\dagger \hat{a}_k \rangle \phi_n^*(\omega_s)\phi_k(\omega_s) \quad (A6)$$
$$= \sum_n^\infty v_n^2 |\phi_n(\omega_s)|^2,$$

and similar for the idler. In the next step, we are now calculating the cross-correlation function

$$G_{si}^{(2)}(\omega_s,\omega_i) = \langle \hat{E}_s^{(-)}(\omega_s)\hat{E}_i^{(-)}(\omega_i)\hat{E}_s^{(+)}(\omega_s)\hat{E}_i^{(+)}(\omega_i)\rangle. \quad (A7)$$

Substituting the fields leads to

$$G_{si}^{(2)}(\omega_s,\omega_i) = \left|\sum_n^\infty v_n u_n \phi_n(\omega_s)\chi_n(\omega_i)\right|^2 + \sum_{n,m}^\infty v_n^2 v_m^2 |\phi_n(\omega_s)|^2 |\chi_m(\omega_i)|^2. \quad (A8)$$

This can be transformed to the final form:

$$G_{si}^{(2)}(\omega_s,\omega_i) = \left|\sum_n^\infty v_n u_n \phi_n(\omega_s)\chi_n(\omega_i)\right|^2 + \langle N(\omega_s)\rangle\langle N(\omega_i)\rangle. \quad (A9)$$

The covariance at high photon numbers expressed in terms of $G_{si}^{(2)}$ is $\text{Cov}[N(\omega_s), N(\omega_i)] = G_{si}^{(2)} - \langle N(\omega_s)\rangle\langle N(\omega_i)\rangle$, so that the final expression for the covariance reads:

$$\text{Cov}[N(\omega_s), N(\omega_i)] = \left|\sum_n^\infty v_n u_n \phi_n(\omega_s)\chi_n(\omega_i)\right|^2. \quad (A10)$$

From this, it is clear that the covariance has to be positive and the shape of the TFS modes can be accessed by measuring the photon number covariance. The measured negativity in Fig. A2 is related to pump depletion, which is not considered here.

## Appendix B: Comparison to simulation

Fig. A1 shows the comparison between the generalized nonlinear Schrödinger equation (GNLSE) simulation and the actual measurement, with $C_{ij}$ calculated for the complete spectrum. The fiber has a core diameter of 18.5 μm (flat-to-flat) and a core wall thickness of 240 nm, these parameters allow one to find the dispersion profile for the simulation from [44] (s-parameter = 0.03). We model the vacuum fluctuations in the GNLSE with the standard approach of one photon per spectral bin with random phase [45]. One can see that the simulation and experiment are in very good agreement. The pump (P) is filtered out in the experiment, since it leads to the saturation of the spectrometer. The measured spectra are a bit narrower than the simulated ones. On the signal side this is a result of the smaller dynamic range, whereas on the idler side the spectrum gets cut at around 286 THz, corresponding to ~1050 nm, which is the sensitivity limit of the silicon chip in the spectrometer. Due to the symmetry of the correlation coefficient we can also see the mirror image of the cross-correlation signal ($CC_M$). In the main text we concentrate only on the cross-correlation signal (CC), which is marked as region of interest (ROI) in Fig. A1. Looking at the cross-correlation, we can clearly notice that the low idler frequencies are in this case correlated with the high signal frequencies.

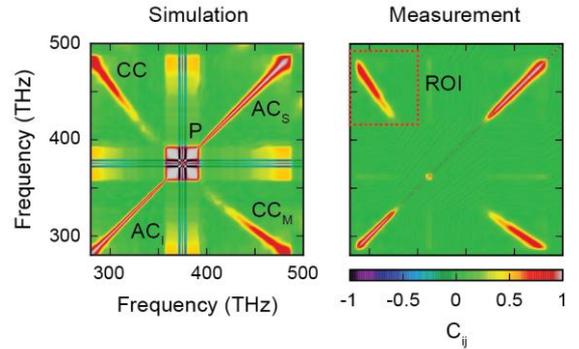

FIG. A1. (Color online) Correlation coefficient for a 30 cm long PCF filled with 76 bar Ar gas, pumped with unchirped 300 fs pulses with an energy of 220 nJ. $AC_{S,I}$ – Auto-correlation signal or idler, CC – Cross-correlation, $CC_M$ – Mirrored cross-correlation, P – Pump, ROI – Region of interest. In all other figures in the paper we show only the ROI, corresponding to the CC signal.

## Appendix C: Power-dependent spectral correlations

Fig. A2 shows the power dependent evolution of the frequency correlations. Here, we did not filter out the pump light. Consequently, because of the limited dynamic range of the spectrometer, we can only see the sidebands appearing at rather high gain values (G = 11.1 and higher). The studied system is a 1 m long kagomé-PCF filled with 76 bar argon gas and pumped with unchirped 280 fs long pulses (3.1 nm bandpass filter). The figure corresponds to curve (II) in Fig. 7(a). In the first two panels of the top row (G = 9.4, 10.3) one can only see the SPM broadened pump. The characteristic shape arises from spectral jittering due to the fluctuations of the pump power [3]. At G = 11.1 the MI sidebands appear. Their cross-correlation shape is distorted due to strong SPM in the system (compare Fig. 5(a)). With increasing pump power, we can observe pump depletion starting at G = 12.0 and getting stronger at G = 12.8. It can be identified as the intensity anti-correlation (dark blue) areas. Also, we see that the SPM loses its correlation. This is due to the overlap of the sidebands with the pump, which cannot be seen in the figure because it is outside the dynamic range of this measurement. Finally, almost all correlations, except the trivial autocorrelations, are lost when the gain is further increased. This corresponds to the case where a broad supercontinuum is generated.

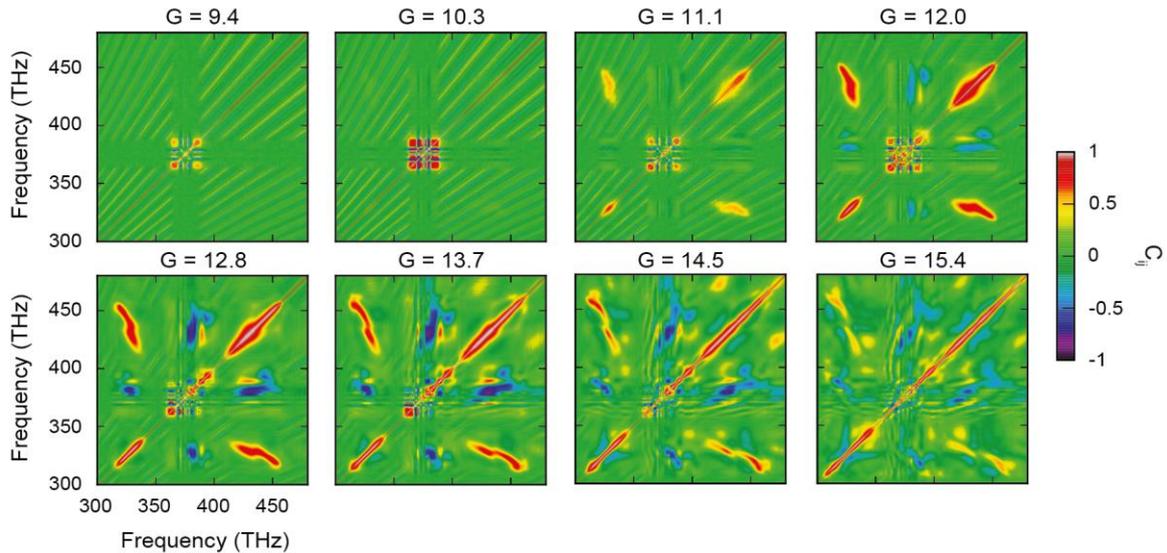

FIG. A2. (Color online) Gain (pump power) dependence of the whole spectrum including the pump in a 1 m long PCF filled with 76 bar argon. This measurement corresponds to curve (II) in Fig. 7(a). It can be seen how correlated MI sidebands appear and lose their correlation again at high pump powers. The diagonal stripes in the top-row figures are caused by the electronic noise of the spectrometer.